\documentstyle[]{article}
\title{ Statistics of skyrmions and the $\nu=5/2$ puzzle }
\author{Jacek Dziarmaga  \\
        Department of Mathematical Sciences,\\
        University of Durham, South Road, Durham, DH1 3LE,\\ 
        United Kingdom.\\
        E-mail address: J.P.Dziarmaga@durham.ac.uk }
\date{November 20, 1996}
\begin{document}
\maketitle
\begin{abstract}

   1) For the hard core interaction there is some freedom left in the
choice of the exact multiskyrmionic wave function's topology. The
statistics of textured quasiholes, analyzed by calculation of the Berry
phase, depends on this choice of topology. 

   2) We find a class of textured two-hole eigenstates of the Coulomb
interaction. There is no definite quantum statistics but there is a
definite rule of how to construct Coulomb eigenstates out of the hard core
wave functions. 

   3) A wave function for the $5/2$ state is constructed according to this
rule. 

\end{abstract}

   Since the experiments on the spin unpolarized FQHE \cite{knight} there
has been considerable interest in skyrmions or textured quasiparticles
\cite{sondhi,fertig,moon}. The theory of skyrmions describes
quantitatively the spin depolarization as the filling fraction is driven
away from $\nu=1$, see \cite{crystal}. One would also like to be able to
construct the whole hierarchy \cite{hh} of FQHE states at other filling
factors and at the same time to predict their spin polarization. The
quantum statistics of skyrmions is the essential ingredient of this
construction.  If the relevant quasiparticles are fermions the hierarchy
construction can explain only the odd denominator states. There is,
however, the remarkable unpolarized state at $\nu=5/2$. In this paper we
address the question of skyrmion statistics and at the same time suggest a
possible explanation of this exceptional even denominator state.

\section{ Statistics of skyrmions for the hard core potential.}

  The wave function of a single skyrmion (textured hole) can be described
by its polynomial part \cite{mfb}

  \begin{equation}\label{hc.20}
  \prod_{k=R+1}^{\infty} z_{k} \prod_{m>n=1}^{\infty}(z_{m}-z_{n}) \;
  |\uparrow_{1}...\uparrow_{R}\downarrow_{R+1}...> \;\;,
  \end{equation} 

where $z=x-iy$ and $R$ is the number of reversed spins. The simplicity of
the hard core potential, $V(z_{1}-z_{2})=V_{0}\;\delta(z_{1}-z_{2})$,
enables an explicit construction of zero energy multiskyrmionic
eigenstates. The quantum statistics can be read from the Berry phase
picked up by these eigenstates under adiabatic exchange of two skyrmions
\cite{asw}. 

  Let us consider the following two-skyrmion eigenstate \cite{st1}

  \begin{equation}\label{hc30} 
  G_{D}(\{z_{k}\},w_{1},w_{2})=
  \prod_{k=R+1}^{\infty}(z_{k}-w_{1})(z_{k}-w_{2})
  \prod_{m>n=1}^{\infty}(z_{m}-z_{n}) \;
  |\uparrow_{1}...\uparrow_{R}\downarrow_{R+1}...>
  \;\;. 
  \end{equation} 

There are two skyrmions localized at $w_{1}$ and $w_{2}$. The Berry phase
picked up by the wave function (\ref{hc30}) during an anticlockwise
exchange of the two skyrmions is $2\pi(1+\frac{R}{2})$. The quantum
statistics depends on the number of reversed spins. 

  An alternative exact multiskyrmionic wave function has been proposed in
\cite{sy},

  \begin{eqnarray}\label{hc40}
  G_{SY}(\{z_{k}\},w_{1},w_{2})&=&
  \prod_{k=1}^{R} [ (z_{k}-w_{1})+(z_{k}-w_{2}) ]
  \prod_{l=R+1}^{\infty}(z_{l}-w_{1})(z_{l}-w_{2})  \nonumber \\
  &&\prod_{m>n=1}^{\infty}(z_{m}-z_{n})
  |\uparrow_{1}...\uparrow_{R}\downarrow_{R+1}...>\;\;. 
  \end{eqnarray}

The statistics of skyrmions in this state turns out to be fermionic like
for the polarized holes. 

  The two considered wave functions are equally good as they are both
exact zero energy eigenstates of the hard core model. Neither the wave
function (\ref{hc30}) nor (\ref{hc40}) is an eigenstate of the Coulomb
interaction, however. The charged skyrmions localized at some definite
$w_{1}$ and $w_{2}$ would interact through the long range electric field,
which, thanks to the Magnus force \cite{stone}, would force them to rotate
one around another with angular velocity dependent on their mutual
distance. The analysis of the skyrmion statistics in terms of the
variational wave functions \cite{sy,st1} can be made rigorous only for the
hard core interaction. As we could see the conclusion depends on the
choice of the topology of the variational wave function. In the following
we are going to use both (\ref{hc30}) and (\ref{hc40}) as generating
functions for Coulomb eigenstates.

\section{Coulomb interaction and the statistics of skyrmions.}

  Let us consider first two polarized electrons confined to the lowest
Landau level (LLL) and interacting through the Coulomb potential. A given
total angular momentum $L<0$ can be split into relative and center of mass
contributions, $L=L_{CM}+L_{rel}$,

  \begin{equation}\label{c1}
  (z_{1}+z_{2})^{|L_{CM}|}(z_{1}-z_{2})^{|L_{rel}|}  \;\;,
  \end{equation}

Both $L_{CM}$ and $L_{rel}$ are good quantum numbers. As $|L_{rel}|$ sets
the distance between electrons, the Coulomb energy is minimized for the
maximal possible $|L_{rel}|$. The fermionic statistics of electrons
constrains $L_{rel}$ to odd integers. The lowest energy eigenstates in the
sectors $L=-1$ and $L=-2$ have the same energy as they both have
$L_{rel}=-1$. In the subspaces $L=-3,-4$ the lowest states are those of
$L_{rel}=-3$. If we define $E(L)$ as the energy of the lowest state in the
subspace of angular momentum $L$, this energy will be degenerate for the
pairs $L=(-1,-2),(-3,-4),(-5,-6)...$ . For bosons the degenerate pairs
would be $L=(-2,-3),(-4,-5),(-6,-7)...$ . The degeneracy between the pairs
is removed by the repulsive interaction. It would not be removed by the
hard core interaction, so that again we would not be able to distinguish
between bosons and fermions. 

   The same idea can be applied to a pair of polarized holes near $\nu=1$. 
In this case it is possible to perform an exact diagonalization of the
Coulomb interaction projected on the LLL in the planar geometry. We use
the one particle orbitals $z^{k} \exp( -\frac{\mid z\mid^{2}}{4} )$,
$k=0,1,2,...$, with corresponding annihilation operators $b_{l}$. The
subspace with a definite angular momentum $L>0$ is spanned by the two hole
states $| h_{1},h_{2}> \equiv \prod_{k\neq h_{1},h_{2}}
b_{k}^{\dagger}|0>$, such that $L=h_{1}+h_{2}$. The energies $E_{pol}(L)$
of the lowest eigenstates are listed in the second row of the Table 1. The
way in which $E_{pol}(L)$ depends on the total angular momentum shows that
the polarized holes are fermions. 

   The two hole Coulomb eigenstates, obtained from the exact
diagonalization, can be constructed from the following generating
function, which is a two hole eigenstate of the hard core interaction,

  \begin{equation}\label{c2}
  G_{pol}(\{z_{k}\},w_{1},w_{2})=
  \prod_{k=1}^{\infty}(z_{k}-w_{1})(z_{k}-w_{2})
  \prod_{m>n=1}^{\infty}(z_{m}-z_{n}) 
  \;|\downarrow_{1}\downarrow_{2}...> \;\;, 
  \end{equation}

with the help of the projection

  \begin{eqnarray}\label{c3}
  &&\psi_{ L_{CM}, L_{rel} }( \{z_{k}\} )= \nonumber \\
  &&\int d^{2}w_{1}\;d^{2}w_{2}\; \exp(-\frac{|w_{1}|^{2}+|w_{2}|^{2}}{4})\; 
  (\bar{w}_{1}+\bar{w}_{2})^{L_{CM}}\;
  (\bar{w}_{1}-\bar{w}_{2})^{L_{rel}}\;
  (w_{1}-w_{2})\;
  G_{pol}(\{z_{k}\},w_{1},w_{2}) \nonumber \\
  &&\equiv
  \lim_{w_{1},w_{2}\rightarrow 0}\;
  \frac{\partial^{L_{CM}}}{\partial(w_{1}+w_{2})^{L_{CM}}}\;
  \frac{\partial^{(L_{rel}-1)}}{\partial(w_{1}-w_{2})^{(L_{rel}-1)}}\;
  G_{pol}(\{z_{k}\},w_{1},w_{2}) \;\;,
  \end{eqnarray}

where equalities are to be understood up to normalization factors. For
this wave function to be nonzero, $L_{rel}$ has to be odd. The idea of
this construction is that (\ref{c2}) is not an eigenstate of the angular
momentum but it is a combination of the angular momentum eigenstates.
Different angular momentum eigenstates are degenerate for the hard core
interaction but their degeneracy is removed by the Coulomb potential. To
obtain a Coulomb eigenstate one has to project (\ref{c2}) on an angular
momentum eigenstate. The relative and CM angular momenta are well defined
because the generating function (\ref{c2}) is symmetric in $w$'s and
homogeneous under simultaneous rescaling of $w$'s and $z$'s,
$(w_{\alpha},z_{k})\rightarrow(\lambda w_{\alpha}, \lambda z_{k})$. 

   After this encouraging exercise we have considered spin textured states
with two holes and one reversed spin. The subspace of angular momentum $L$
is spanned by the states $|h_{1},h_{2},h_{3};s>=
 a_{s}^{\dagger}\prod_{l\neq h_{1},h_{2},h_{3}}b_{l}^{\dagger}|0>$, where
$a_{s}$ annihilates a spin up electron in the $s$-th orbital and
$L=h_{1}+h_{2}+h_{3}-s$. For any $L$ the dimension of the corresponding
Hilbert space in infinite. We had to truncate the Hilbert space by
assuming that the orbitals higher than some cut-off $M$ were not excited,
$0\leq h_{1},h_{2},h_{3},s \leq M$. Outside the ring of the $M$-th orbital
the state was effectively forced to be ferromagnetic.  The spin texture of
the skyrmionic states like (\ref{hc.20}) is localized just in a power law
way. The skyrmionic tails can be expected to be distorted by the imposed
cut-off. The quantities, like the energy of an unpolarized ground state,
approach their asymptotic values in an algebraic way. In this situation,
for any $L$, we have repeated the calculations for a range of around $15$
values of M. Then the energies were extrapolated to $1/M=0$ by a fit with
a rational function of $1/M$.

\subsection{Lowest Landau level}

   Similarly as for the polarized states, one can use the generating
functions (\ref{hc30},\ref{hc40},\ref{c2}) to construct the Coulomb 
eigenstates

  \begin{equation}\label{c4}
  \lim_{w_{1},w_{2}\rightarrow 0}\;
  \frac{\partial^{A}}{\partial(w_{1}+w_{2})^{A}}\; 
  \frac{\partial^{B}}{\partial(w_{1}-w_{2})^{B}}\;
  G(\{z_{k}\},w_{1},w_{2}) \;\;.
  \end{equation}

The lowest state for $L=1$ is polarized. It can be constructed with
$G=S^{+}G_{pol}$ and $A=B=0$, 

  \begin{eqnarray}\label{c.20}
  \prod_{k=1}^{\infty}z_{k}^{2}\prod_{m>n=1}^{\infty}(z_{m}-z_{n})
  |\uparrow_{1}\downarrow_{2}...>\;\;.
  \end{eqnarray}

   In the second quantization language this state reads $\sum_{a=2}^{M}
|0,1,a;a>$, where $M$ is the cut-off. The overlap with the ground state is
$0.904$. The influence of the cut-off on the polarized states is the
strongest. In the exact state the reversed spin would be uniformly
distributed over the plane but the cut-off forces it to be localized
around the origin. The contribution from the states $|0,1,a;a>$ to the
norm squared of the ground state is $0.993$. Also the extrapolated energy
matches well with the exact energy of the polarized $L=1$ state, see Table
1. 

    For $L=3$ we get the ground state from $G=G_{D}$ with $A=B=0$,
 
  \begin{equation}\label{c.10}
  \prod_{k=2}^{\infty} z_{k}^{2}
  \prod_{m>n=1}^{\infty}(z_{m}-z_{n})\;
  |\uparrow_{1}\downarrow_{2}...> \;\;.
  \end{equation}
 
   The second quantization expression is $\sum_{a=0}^{M-2}
\frac{(-1)^{a}}{\sqrt{(a+1)(a+2)}}
 |0,1,a+2;a>$. The overlap with the ground state is $0.977$ and the states
$|0,1,a+2;a>$ contribute $0.999$ of its norm squared. 

   The $L=5$ ground state can be obtained again from $G=G_{D}$ but this
time $A=0,B=2$. The overlap is $0.93$ and the contribution to the norm
squared amounts to $0.98$. 

   The $L=2,4$ states can be obtained from $G=G_{SY}$. For $L=2$ we have
to take $A=B=0$,

  \begin{equation}\label{c.30}
  z_{1}
  \prod_{k=2}^{\infty} z_{l}^{2}
  \prod_{m>n=1}^{\infty}(z_{m}-z_{n})\;
  |\uparrow_{1}\downarrow_{2}...> \;\;.
  \end{equation}
  
The second quantization form of this state is
$\sum_{a=0}^{M-1}\frac{(-1)^{a}}{\sqrt{a+1}}|0,1,a+1;a>$. The overlap with
the numerical ground state is $0.95$ and the contribution of the states
$|0,1,a+1;a>$ to its norm squared is $0.97$. We observe some contribution
to the tail of the ground state from the family of states $|1,2,a;a+1>$
with $a=3,4...$ but the core remains undistorted. 

   The $L=4$ ground state is obtained with $A=0,B=2$. The overlap is
$0.946$ and the contribution to the norm squared is $0.99$. 

   We find the data for $L>5$ inconclusive. The states with higher $L$ are
more sensitive to the imposed cut-off. On the basis of the data obtained
so far we can conclude that, except the polarized $L=1$ state, the $L$-odd
states are generated from $G_{D}$ with increasing $B$, $L=3+B$. The
$L$-even states are generated from $G_{SY}$ so that $L=2+B$. In any case
$A=0$. It is energetically favorable to choose the right generating
function $G$ so that for a given angular momentum $L$ one can keep the
center of mass angular momentum $A=0$ and increase the relative angular
momentum by an appropriate even $B$. 

   The third row of the Table 1 gives the extrapolated energies of the
ground states for various $L$. There are no characteristic steps, which
could define the quantum statistics. This lack of degeneracy is not just a
numerical artifact as we can see from the insight into the nature of the
ground states for various $L$. Like for the hard core potential the
skyrmions are neither bosons nor fermions. 

   The constructive message is that the Coulomb eigenstates can be
constructed from the hard core eigenstates by appropriate projections.
Both $G_{D}$ and $G_{SY}$ are equally good for this task. If we restricted
to just one of the generating functions, it would give rise to a staircase
of characteristic steps. $G_{D}$ would give steps $L=(3,4),(5,6)...$ and
$G_{SY}$ would give degenerate ground states for $L=(1,2),(3,4)...$.
However, the two have to compete for any $L\geq 3$. According to the data
obtained so far $G_{D}$ wins for $L$-odd and $G_{SY}$ wins for $L$-even.
The generating function for a given $L$ has to be chosen so that to make
the best possible use of the relative angular momentum $B$ and keep $A$ as
close to $0$ as possible.

\subsection{$n=1$ Landau level}

   We have performed similar calculations in the second Landau level. We
confirm \cite{shll} that there are no single skyrmions. The charge of a
single hole is not sufficient to overcome the increased spin stiffness. We
however find some unpolarized states in the two hole case. 

   The $L$-odd states are found to be polarized. The contribution of the
$|0,1,a;a>$ states to the norm squared of the $L=1$ ground state is
$0.9993$. Analogous contributions for the $L=3,5$ states are $0.993$ and
$0.985$ respectively. The extrapolated energies of these states match the
exact energies of the polarized ground states, see Table 2. 

   The $L=2$ state is, similarly as in the lowest Landau level, obtained
from $G_{SY}$ with $A=B=0$. The overlap with the ground state and the
contribution to its norm squared are $0.87$ and $0.78$ respectively. The
extrapolated energy is lower than the exact energy of the polarized $L=2$
state. The energy difference is around $5\%$ of the gross energy of the
polarized hole. The state should become more stable than its polarized
counterpart below the magnetic field of the order of $20;T$. 

   The overlap of the $L=4$ ground state with the $G=G_{SY}$ and $A=0,B=2$
state is $0.71$ and the contribution to the norm squared is $0.59$. On the
other hand the contribution from the families of polarized states
$|0,4,a;a>$ and $|1,3,a;a>$ is just $0.506$ so the state can hardly be
believed to be polarized. The extrapolated energy is close to the energy
of the polarized state. This state is likely to be stable only at
extremely low magnetic fields. 

   The data for higher $L$ are not convincing. In general it is observed
that the contribution from the polarized states to the $L$-even states is
lower (less than $90\%$ and decreasing with the cut-off) than their
contribution to $L$-odd states (more than $97\%$ and increasing). This
regularity may suggest that also for higher $L$ the $L$-odd states are
polarized and the $L$-even states are generated from $G_{SY}$. The set of
generating functions in the second Landau level is restricted to the
functions $S^{+}G_{pol},G_{SY}$.

\section{The $\nu=5/2$ unpolarized state.}

   We interpret the $5/2$ state as the $1/2$ unpolarized state in the
$n=1$ Landau level. At the filling factor of $1/2$ the expectation value
of the angular momentum per pair of holes is $2$. If we were to construct
a droplet of such state in the planar geometry, we would have to find the
ground state in the $L=N(N-1)$ subspace of the $N$ holes' Hilbert space.
For an unpolarized $1/2$ state we have to reverse one spin per two holes.
Let us try to construct a state of 4 textured holes from a generalization
of the $G_{SY}$ function (\ref{hc40})

  \begin{eqnarray}\label{x10}
  &&\psi_{4\;holes}(\{z_{k}\})= \nonumber \\
  &&\lim_{w_{1},..,w_{4}\rightarrow 0}\;
  (\partial_{w_{1}}+...+\partial_{w_{4}})^{A}\;
  \prod_{\beta>\alpha=1}^{4}
  (\partial_{w_{\alpha}}-\partial_{w_{\beta}})^{ B_{\alpha\beta}+1 }\;
  (w_{\alpha}-w_{\beta})^{4}
  \nonumber \\
  && 
  \prod_{k=1}^{2}[\sum_{\gamma=1}^{4}
                  \prod_{\delta\neq\gamma}(z_{k}-w_{\delta})]\;
  \prod_{l=3}^{\infty}\prod_{\sigma=1}^{4}(z_{l}-w_{\sigma})
  \prod_{m>n=1}^{\infty}(z_{n}-z_{m})\;
  |\uparrow_{1}\uparrow_{2}\downarrow_{3}...> \;\;.
  \end{eqnarray}

There are six pairs of holes so the total angular momentum should be
chosen as $L=12$. To minimize the Coulomb energy as much as possible we
set $L_{CM}=A=0$, imposing effectively the constraint $w_{1}+..+w_{4}=0$
on the generating function. With this constraint $L=8$ for vanishing
$B$'s. For the angular momentum to be $L=12$, the six $B$'s must satisfy
the constraint $\sum_{\beta>\alpha=1}^{4} B_{\alpha\beta}=4$. At least two
of them have to vanish, say $B_{13}=B_{24}=0$. In this way the generating
function can be further restricted by $w_{3}=w_{1}$ and $w_{4}=w_{2}$. The
holes are grouped into textured pairs $(1,3)$ and $(2,4)$. The formula
(\ref{x10}) simplifies to

  \begin{eqnarray}\label{x20}
  &&\psi_{2\;pairs}(\{z_{k}\})=   \nonumber \\
  &&\lim_{w\rightarrow 0}\;
  \frac{ \partial^4 }{ \partial w^{4} }
  \prod_{k=1}^{2}[ (z_{k}-w)(z_{k}+w)^{2}
                  +(z_{k}+w)(z_{k}-w)^{2} ]\;
  \prod_{l=3}^{\infty} (z_{l}-w)^{2}(z_{l}+w)^{2}  \nonumber \\
  &&\prod_{m>n=1}^{\infty}(z_{n}-z_{m})\;
  |\uparrow_{1}\uparrow_{2}\downarrow_{3}...> \;\;,
  \end{eqnarray} 

where $w=w_{1}-w_{2}$ is the relative coordinate of the two pairs.  The
above construction can be repeated for any even number of holes $2R$ at
the angular momentum $L=2R(2R-1)$. The resulting wave function is

  \begin{eqnarray}\label{x30}
  &&\psi_{R\;pairs}(\{z_{k}\})=  
  \lim_{w_{1},..,w_{R}\rightarrow 0}\;
  \prod_{\beta>\alpha=1}^{R}
  (\partial_{w_{\alpha}}-\partial_{w_{\beta}})^{8} 
  (w_{\alpha}-w_{\beta})^{4}                      \nonumber \\
  && \prod_{k=1}^{R}[\sum_{\gamma=1}^{R}(z_{k}-w_{\gamma})
                    \prod_{\delta\neq\gamma}(z_{k}-w_{\delta})^{2}]\;
  \prod_{l=R+1}^{\infty}\prod_{\sigma=1}^{R}(z_{l}-w_{\sigma})^{2}
                                                   \nonumber \\
  &&\prod_{m>n=1}^{\infty}(z_{n}-z_{m})\;
  |\uparrow_{1}..\uparrow_{R}\downarrow_{R+1}...> \;\;,
  \end{eqnarray}

where $w$'s are pairs' coordinates. As we could see, when we assume the
density of holes to be the same as in the $1/2$ state, the textured holes
form unpolarized pairs first and then the pairs condense into a Laughlin
state at the bosonic filling factor $\nu_{b}=1/8$. The interpretation in
terms of bosonic pairs makes sense, if we restrict to the subspace of
$G_{SY}$ wave functions. We have chosen $G_{SY}$ as a generating function
because it enables construction of the $1/2$ state with $A=0$ for any $R$.
Also the distribution of $B$'s is uniform like in the Laughlin state. 
From the Subsection 2.2 we know that pairs are likely to be stable even in
quite strong magnetic fields.

\section{Spin of the wave functions}

   The wave functions considered in this paper have definite $S_{z}$ by
construction but it is not obvious what is their total spin $S$. In this
Section we will consider various cases in order of increasing difficulty. 
The argument is generalization of an analogous proof in Ref.\cite{mfb}.

\subsection{Spin of the $G_{SY}$-generated two skyrmion states}

    Let us consider first the case of the states obtained from the
generating function (\ref{hc40}) with the projection (\ref{c4}). The
states $(A,B)$ with nonzero $A$ can be obtained from the $(0,B)$ states
just by increasing the center of mass angular momentum with the operator
$(\partial_{z_{1}}+\partial_{z_{2}}+...+\partial_{z_{N}})$.  This
operation does not change the spin quantum numbers, so we can restrict to
the $(0,B)$ states

  \begin{equation}\label{s10}
  \lim_{w^{2}\rightarrow 0}
  \frac{\partial^{b}}{\partial (w^{2})^{b} }
  \prod_{k=1}^{R} z_{k}
  \prod_{l=R+1}^{\infty}(z_{l}^{2}-w^{2}) 
  \prod_{m>n=1}^{\infty}(z_{m}-z_{n})\;
  |\uparrow_{1}...\uparrow_{R}\downarrow_{R+1}...>\;\;,
  \end{equation}

where $w=w_{1}-w_{2}$ is the relative complex coordinate and $B=2b$. For 
any wave
function $\psi$ its "bosonic" part $\psi_{B}$ can be defined by $\psi=
\psi_{B} \prod_{m>n=1}^{\infty}(z_{m}-z_{n})$. $\psi$ has the same spin
quantum numbers as $\psi_{B}$. The bosonic part of (\ref{s10}) is

  \begin{equation}\label{s20}
  \lim_{w^{2}\rightarrow 0}
  \frac{\partial^{b}}{\partial (w^{2})^{b} }  
  \prod_{k=1}^{R} z_{k}
  \prod_{l=R+1}^{N}(z_{l}^{2}-w^{2})\;
  |\uparrow_{1}...\uparrow_{R}\downarrow_{R+1}...>\;\;.
  \end{equation}
                    
According to our usual convention the exponential factors are neglected
and we display only one spinor component - the one with the spins
$(1,2,...,R)$ pointing up and the rest pointing down. As a regulator we
keep the number of electrons $N$ finite. The number of spinor components
is $(\stackrel{N}{R})$. 

  Let us find out what is the expectation value of the operator 

  \begin{equation}\label{s30}
  S^{2}=S_{z}^{2}+\frac{S_{+}S_{-}+S_{-}S_{+}}{2}=
        (\frac{N}{2}-R)^2+\frac{N}{2}+
        2\sum_{k>l=1}^{N} s_{+}^{(l)}s_{-}^{(k)} \;\;
  \end{equation}

in the state (\ref{s20}). $s$'s are single electron spin operators and the
indices $k,l$ run over electrons. The last part of (\ref{s30}) does not
contribute to the expectation value of $S^{2}$ in the state (\ref{s20}). 
The $"s_{+}s_{-}"$ part mixes different spinor components. The spinor
components for two different groups of up-spins are orthogonal. This
property easily follows from the fact that in the wave function
(\ref{s20}) the coordinates of spin-up electrons appear with first power,
while the coordinates of the spin-down electrons appear with even powers.
Thus the expectation value of $S^{2}$ is just
$(\frac{N}{2}-R)^2+\frac{N}{2}$, what implies that in the limit of large
$N$ the spin tends to $S=\frac{N}{2}-R$. 

  We can conclude that for the states (\ref{s10}) the total spin is
$S=\frac{N}{2}-R$, which is the lowest possible value for the given
$S_{z}=-\frac{N}{2}+R$.

\subsection{Spin of the $G_{D}$-generated two skyrmion states}

   Similarly as in the subsection 4.1 it is enough to consider the states
generated (\ref{hc30},\ref{c4}) with $A=0$. Their bosonic part is

   \begin{eqnarray}\label{w10}
   && C^{'}\lim_{w^{2}\rightarrow 0}
   \frac{\partial^{b}}{\partial(w^{2})^{b}}
   \prod_{k=R+1}^{N}(z_{k}^{2}-w^{2})\;
   |\uparrow_{1}...\uparrow_{R}\downarrow_{R+1}...\downarrow_{N}>=
   \nonumber \\
   && C \sum_{ \stackrel{ k_{1}<...<k_{b} }
                       { k_{1},...,k_{b}\in\{R+1,...,N\} } }
   \prod_{ \stackrel{k=R+1,...,N}
                    {k\neq k_{i}} }
   z_{k}^{2}\; 
   |\uparrow_{1}...\uparrow_{R}\downarrow_{R+1}...\downarrow_{N}>\equiv
   \nonumber \\
   && \psi_{\{1,...,R\}}\;
   |\uparrow_{1}...\uparrow_{R}\downarrow_{R+1}...\downarrow_{N}> \;\;.
   \end{eqnarray}

$C$ and $C^{'}$ are normalization constants such that the spinor component
$\psi_{\{1,...,R\}}$ for the spins $\{1,...,R\}$ pointing up, like any
other spinor component, is normalized to unity, 

   \begin{equation}\label{w20}
   C^{2}=\frac{1}{ (\stackrel{N-R}{b}) (16\pi)^{(N-R-b)} } \;\;.
   \end{equation}

The expectation value of the operator $S^{2}$ (\ref{s30}) in the state
(\ref{w10}) is given by 

   \begin{equation}\label{w30}
   <S^{2}>=(\frac{N}{2}-R)^2+\frac{N}{2}+
           R(N-R)
           < \psi_{\{1,...,R\}} |
             \psi_{\{2,...,R+1\}} > \;\;.
   \end{equation}

Unlike in the case of $G_{SY}$-generated states different spinor
components are not orthogonal in general. The overlap can be worked
out as
 
   \begin{eqnarray}\label{w40}
   && < \psi_{ \{1,...,R\} } | \psi_{ \{2,...,R+1\} } >=  \nonumber \\
   && C^{2} 
   < \sum_{ R<k_{1}<...<k_{b}\leq N } 
     \prod_{ \stackrel{ k=R+1,...,N }{ k\neq k_{i} } } z_{k}^{2} \; | 
     \sum_{ \stackrel{ l_{1}<...<l_{b} }
                     { l_{i}\in\{1,R+2,...,N } } 
     \prod_{ \stackrel{ l=1,R+2,...,N }
                      { l\neq l_{i} } } z_{l}^{2} \;> =   \nonumber \\
   && C^{2} \sum_{ \stackrel{ k_{2}<...<k_{b} }
                            { k_{i}\in\{R+2,...,N\} } }
            \sum_{ \stackrel{ l_{2}<...<l_{b} }
                            { l_{i}\in\{R+2,...,N\} } }
            < \prod_{ \stackrel{k=R+2,...,N}{k\neq k_{i}} } z_{k}^{2} \; |
              \prod_{ \stackrel{l=R+2,...,N}{l\neq l_{i}} } z_{l}^{2} \;>=
                                                          \nonumber \\
   && C^{2} (\stackrel{N-R-1}{b-1})
            < \prod_{k=R+b+1}^{N} z_{k}^{2} \; | \;
              \prod_{l=R+b+1}^{N} z_{l}^{2} > =           \nonumber \\
   && \frac{b}{16\pi(N-R)}  \;\;.                              
   \end{eqnarray}

One can easily find out, when this result is substituted to
Eq.(\ref{w30}), that for large $N$ the spin tends to $S=\frac{N}{2}-R$, no
matter what is $b$. We can conclude that also the $G_{D}$-generated two
skyrmion states have the lowest possible spin of $\frac{N}{2}-R$.

\subsection{Spin of the $5/2$ droplet state}

  The state (\ref{x30}) is a droplet of the $\nu=1/2$ unpolarized state in
the sea of polarized $\nu=1$ state. The $(1,2,...,R)$ spins up spinor
component of its bosonic part is

  \begin{eqnarray}\label{y10}
  && C \;\lim_{w_{1},..,w_{R}\rightarrow 0}\;
  \prod_{\beta>\alpha=1}^{R} 
  (\partial_{w_{\alpha}}-\partial_{w_{\beta}})^{8}
  (w_{\alpha}-w_{\beta})^{4}
  \nonumber \\
  &&\prod_{k=1}^{R}[\sum_{\gamma=1}^{R}(z_{k}-w_{\gamma})
                    \prod_{\delta\neq\gamma}(z_{k}-w_{\delta})^{2}]\;
  \prod_{l=R+1}^{N}\prod_{\sigma=1}^{R}(z_{l}-w_{\sigma})^{2}
  |\uparrow_{1}..\uparrow_{R}\downarrow_{R+1}...>\equiv \nonumber \\
  &&\psi_{\{1,...,R\}}
  |\uparrow_{1}..\uparrow_{R}\downarrow_{R+1}...> \;\;. 
  \end{eqnarray}

$C$ is a normalization factor such that this spinor component is
normalized to unity. 

   As a warm-up exercise let us consider the case of $R=2$, when
everything is fairly explicit,

  \begin{eqnarray}\label{y20}
  && C\; \lim_{w^{2}\rightarrow 0}\;
  \frac{ \partial^2 }{ \partial (w^{2})^2 }
  \prod_{k=1,2} z_{k}(z_{k}^{2}-w^2) \;
  \prod_{l=3,4,...,N} (z_{l}^{2}-w^{2})^{2} \;
  |\uparrow_{1}\uparrow_{2}\downarrow_{3}...>+ 
  \end{eqnarray} 

The coordinates of spin-down electrons appear with even powers while the
powers of the spin-up electrons' coordinates are odd. Because of this
"odd-even" property the spinor components for different groups of up-spins
are orthogonal. Thus, in a similar way as in the Section 4.1, we find that
the wave function (\ref{y20}) has the lowest possible spin
$S=\frac{N}{2}-2$. 

   Generalization to higher $R$ does not seem to be straightforward. It is 
not obvious if different spinor components are orthogonal. The expectation
value of $S^{2}$ in the state (\ref{y10}) is
 
  \begin{equation}\label{y30}
 <S^{2}>=(\frac{N}{2}-R)^2+\frac{N}{2}+
        R(N-R) 
        < \psi_{\{2,...,R+1\}} |
          \psi_{\{1,3,...,R+1\}} > \;\;,
  \end{equation}
 
where the last contribution comes from the $"s_{+}s_{-}"$ part of the
operator $S^{2}$. To show that the wave function (\ref{y10}) has the
lowest possible spin, it is enought to prove that the overlap
$<\psi_{\{2,...,R+1\}} | \psi_{\{1,3,...,R+1\}} >$ tends to zero in the
limit of large $N$. 

   One can split each spinor component into two parts, say
$\psi_{\{2,...,R+1\}}=\phi_{1}+\phi_{2}$. $\phi_{1}$ is the part in which
$z_{1}$ appears only in the maximal power of $2R$, while $\phi_{2}$
contains all the contributions with lower powers of $z_{1}$, compare with
Eq.(\ref{y10}). The two parts are orthogonal, $<\phi_{1}|\phi_{2}>=0$. In
the Equation (\ref{y10}) the number of derivatives with respect to $w$'s
is finite. With increasing $N$ the contribution from $\phi_{2}$ becomes
negligible as compared to the contribution from $\phi_{1}$. As the sum of
the two orthogonal contributions is normalized to unity, the norm of
$\phi_{2}$ must vanish for $N\rightarrow\infty$. In a similar way we can
split $\psi_{\{1,3,...,R+1\}}=\phi_{3}+\phi_{4}$. $z_{2}$ appears in
$\phi_{3}$ in the maximal power of $2R$ only and $\phi_{4}$ is its
orthogonal complement. It is important to realize that $z_{2}$ appears in
$\psi_{\{2,...,R+1\}}$ and $z_{1}$ appears in $\psi_{\{1,3,...,R+1\}}$
with at most $(2R-1)$ power. Because of that

  \begin{equation}\label{y40}
  < \psi_{\{2,...,R+1\}} | \psi_{\{1,3,...,R+1\}} >=
  <\phi_{2}| \psi_{\{1,3,...,R+1\}} >+
  < \psi_{\{2,...,R+1\}} |\phi_{4}> \;\;.
  \end{equation}

The overlap can be estimated as

   \begin{eqnarray}\label{y50}
  &&|< \psi_{\{2,...,R+1\}} | \psi_{\{1,3,...,R+1\}} >|\leq \nonumber \\
  &&|<\phi_{2}| \psi_{\{1,3,...,R+1\}} >|\;+\;
    |< \psi_{\{2,...,R+1\}} |\phi_{4}>|\leq                 \nonumber \\
  &&||\phi_{2}|| \; + \; ||\phi_{4}|| \;\;.
  \end{eqnarray} 

The last estimate holds thanks to the Cauchy-Schwarz inequality and the
unit normalization of the spinor components. As the norms of $\phi_{2}$
and $\phi_{4}$ vanish for infinite $N$, the overlap has to vanish too. 

   The proposed droplet of the $5/2$ state has the lowest possible spin of
$S=\frac{N}{2}-R$. It is a droplet of unpolarized state.

\section{Remarks}

  The proposed $5/2$ state is due to the condensation of pairs of textured
holes in the second Landau level into the $1/8$ Laughlin state. The
$K$-matrix of such a state should have one negative eigenvalue. The
constructed state has two separated edges. One is the edge of the pair
condesate, where the filling fraction increases from $5/2$ to $3$. The
outer edge is the polarized \cite{edges} edge where the density drops from
$3$ to null. The condesate of $R$ pairs also admits an interpretation as a
spin texture of winding number $2R$. 

  It should be mentioned that the pairing mechanism proposed in this
paper is different from that of Haldane and Rezayi \cite{hare}. In
\cite{hare} the genuine Cooper pairs are stable thanks to short range
attractive interactions (hollow core model). Our pairs are formed of
textured holes and the pair formation is due to distorted ferromagnetic
order.

\paragraph{Acknowledgement.} 

I would like to thank Meik Hellmund for discussions, Roberto Tateo for
sharing his extrapolation program and Rajiv Kamilla for drawing
Ref.\cite{shll} to my attention. This research was supported by UK PPARC.

\subsection*{Table 1}

\begin{tabular}{|c|c|c|c|c|c|c|}
\hline
     L          &      1       &      2      &      3       &     4      &      5     \\ 
\hline
  $E_{pol}(L)$  &   2.35355    &   2.35355   &   2.22097    &   2.22097  &  2.17401   \\
\hline
 $E_{unpol}(L)$ &  2.3538(2)   &  2.3067(4)  &  2.2038(1)   &  2.183(2)  &  2.1409(5) \\
\hline
 $\Delta E(L)$  & -0.0002(2)   & -0.0468(4)  & -0.0171(1)   & -0.038(2)  & -0.0331(5) \\ \hline
\end{tabular}\\

The unit of energy is the gross energy of the polarized hole 
$\varepsilon_{-}=1.25331\frac{e^2}{\kappa l}$. 
The numbers in brackets are extrapolation errors of the last digit.

\subsection*{Table 2}

\begin{tabular}{|c|c|c|c|c|c|c|}  
\hline
     L          &      1       &      2      &      3       &     4      &      5     \\       
\hline
  $E_{pol}(L)$  &   2.441      &   2.441     &   2.335      &   2.335    &  2.247     \\           
\hline
 $E_{unpol}(L)$ &   2.439(3)   &   2.395(3)  &   2.332(3)   &   2.333(4) &  2.242(7)  \\           
\hline
 $\Delta E(L)$  &  -0.002(3)   &  -0.046(3)  &  -0.003(3)   & -0.002(4)  & -0.005(7)  \\    
\hline
\end{tabular}\\
 
The unit of energy is the gross energy of the polarized hole
$\varepsilon_{-}=0.93999\frac{e^2}{\kappa l}$. The numbers in
brackets are extrapolation errors of the last digit.

\end{document}